\documentclass[11pt,twoside]{article}  % Leave intact
\usepackage{apn3conf}

\begin{document}   % Leave intact

%-----------------------------------------------------------------------
%		            Paper Title 
%-----------------------------------------------------------------------
% Enter the title of the paper.
%
% EXAMPLE: \title{A Breakthrough in Astronomical Software Development}
% 
% If your title is so long as to fill the page header when you print it,
% then please supply a short form as a \titlemark.
%
% EXAMPLE: 
%  \title{Rapid Development for Distributed Computing, with Implications
%         for the Virtual Observatory}
%  \titlemark{Rapid Development for Distributed Computing}
%

\title{Outflows and Jets I: a personal take}
%\titlemark{ }

%-----------------------------------------------------------------------
%		          Authors of Paper
%-----------------------------------------------------------------------
% Enter the authors followed by their affiliations.  The \author and
% \affil commands may appear multiple times as necessary (see example
% below).  List each author by giving the first name or initials first
% followed by the last name.  Authors with the same affiliations
% should grouped together. 
%
% EXAMPLE: \author{Margaret Meixner\altaffilmark{1}, Letizia Stanghellini,
%			Howard Bond} 
%          \affil{Space Telescope Science Institute, 
%                 3700 San Martin Dr.,  Baltimore, MD 21218}
%          \author{Joel Kastner}
%          \affil{Rochester Institute of Technology}
%
%          \altaffiltext{1}{Astronomy Department, UIUC}
%
% In this example, the first three authors, "Meixner", "Stanghellini", and
% "Bond" are affiliated with "STScI".  "Meixner" has an alternate 
% affiliation with the "Astronomy Department at UIUC".  The fourth author,
% "Kastner", is affiliated with "Rochester Institute of Technology"

\author{Orsola De Marco}
\affil{American Museum of Natural History}

%-----------------------------------------------------------------------
%			 Contact Information
%-----------------------------------------------------------------------
% This information will not appear in the paper but will be used by
% the editors in case you need to be contacted concerning your
% submission.  Enter your name as the contact along with your email
% address.
% 
% EXAMPLE:  \contact{Dennis Crabtree}
%           \email{crabtree@cfht.hawaii.edu}
%

\contact{Orsola De Marco}
\email{orsola@amnh.org}

%-----------------------------------------------------------------------
%		      Author Index Specification
%-----------------------------------------------------------------------
% Specify how each author name should appear in the author index.  The 
% \paindex{ } should be used to indicate the primary author, and the
% \aindex for all other co-authors.  You MUST use the following
% syntax: 
%
% SYNTAX:  \aindex{LASTNAME, F. M.}
% 
% where F is the first initial and M is the second initial (if
% used).  This guarantees that authors that appear in multiple papers
% will appear only once in the author index.  
%
% EXAMPLE: \paindex{Crabtree, D.}
%          \aindex{Manset, N.}        
%          \aindex{Veillet, C.}        
%
% NOTE: this information is also used to build the author list that
% appears in the table of contents.  Authors will be listed in the order
% of the \paindex and \aindex commmands.
%

\paindex{De Marco, O.}

%-----------------------------------------------------------------------
%		      Author list for page header	
%-----------------------------------------------------------------------
% Please supply a list of author last names for the page header. in
% one of these formats:
%
% EXAMPLES:
% \authormark{LASTNAME}
% \authormark{LASTNAME1 \& LASTNAME2}
% \authormark{LASTNAME1, LASTNAME2, ... \& LASTNAMEn}
% \authormark{LASTNAME et al.}
%
% Use the "et al." form in the case of seven or more authors, or if
% the preferred form is too long to fit in the header.

\authormark{De Marco}

%-----------------------------------------------------------------------
%			Subject Index keywords
%-----------------------------------------------------------------------
% Enter up to 6 keywords describing your paper.  These will NOT be
% printed as part of your paper; however, they will be used to
% generate an object index and a subject index for the proceedings.  
% There is no standard list,  however, individual object names are
% encouraged and one or two word descriptions of the topics (e.g.MHD, 
% ionized gas) are useful. 
%
% EXAMPLE:  \keywords{NGC 7027, AFGL 2688, HD 161796, binary stars,
%                      dust,  molecular gas}
%


%-----------------------------------------------------------------------
%			       Abstract
%-----------------------------------------------------------------------
% Type abstract in the space below.  Consult the User Guide and Latex
% Information file for a list of supported macros (e.g. for typesetting 
% special symbols). Do not leave a blank line between \begin{abstract} 
% and the start of your text.

\begin{abstract}          % Leave intact
As a session chair I have the privilege to write down a few comments on
four talks that were delivered under the general title of ``outflows and jets I".
This and the session that followed it, grouped work on circumstellar 
environments of, post-AGB stars, proto-PNe and PNe.
After summarizing some of the highlights, I pose the question of which observations, if any,
reveal the presence of a common envelope ejection in the past of a PN.
Finally, I point out that the finding that most or all PNe have a binary origin, while
solving some of our problems, might be in contrast with the number of known close main 
sequence binaries.
\end{abstract}

%-----------------------------------------------------------------------
%			      Main Body
%-----------------------------------------------------------------------
% Place the text for the main body of the paper here.  You should use
% the \section command to label the various sections; use of
% \subsection is optional.  Significant words in section titles should
% be capitalized.  Sections and subsections will be numbered
% automatically. 
%
% EXAMPLE:  \section{Introduction}
%           ...
%           \subsection{Our View of the World}
%           ...
%           \section{A New Approach}
%
% It is recommended that you look at the sample papers, sample1.tex
% and sample2.tex, for examples for formatting references, footnotes,
% figures, equations, html links, lists, and other special features.  

{\bf Highlights from the presentations of Sahai, Su, Corradi and Cox.}
From the days of his first HST snapshot survey of young planetary nebulae (PNe) and 
proto-PNe
(PPNe; Sahai \& Trauger 1998), Raghvendra
Sahai has made a very convincing argument that the seed of bipolar symmetry in PN shaping springs
into action very early in the post Asymptotic Giant Branch (AGB) life of a star.
Jet-like outflows suddenly appear during the late-AGB or early post-AGB and 
carve holes into the spherical AGB mass-loss. Any subsequent mass-loss results in PN 
morphologies that bare the imprint of this early hole-carving. 

These young, extremely bipolar PPNe, seen in reflected stellar light or ionized radiation 
are investigated further by Pierre Cox and others in these
proceedings, by looking at molecular light (H$_2$ and CO). Molecular observations can
break the degeneracy of whether the observed morphologies describe ionized 
(optical line radiation) or neutral (optical stellar continuum reflected by dust in the neutral
regions) regions and refine the location and dynamics of the jets-like structures and other 
components of the AGB envelope. Cox and collaborators also remark on the
extreme similarity in the morphologies and kinematics observed across the PPN-PN division,
which points to one jet-shaping phenomenon, acting from very early on, and carrying on
into the PN phase. 

Further out from the inner regions of PPNe and young PNe, we observe
circular concentric rings in reflected light.
In her comprehensive, quantitative study of the rings, Kate Su presents a summary of ring 
properties around 6 PPNe, 5 PNe and one AGB star. Three ring-forming scenarios from the literature
are compared. Binary models, where periodic incursions of a companion
during the AGB phase modulate the mass-loss. A fluid instability model, where the dusty outflow
is unstable in a way that can lead to the observed structures and, finally, a magnetic
activity model, where either magnetic surface spots or magnetic pressure oscillations
are responsible for the mass-loss modulation.

The persistence of the ring structures all the way into the PN phase, appears to indicate 
that, 
after their creation during the AGB phase, the gas that bares the ring signature remains relatively 
undisturbed. This reinforces the theory that the shaping mechanism acts over small solid 
angles, in agreement with the
Sahai and Trauger (1998) scenario of hole-piercing by jet-like
structures in the early PPN phase. 

I expect that in the near future, this excellent quantitative work will be able to 
rule out several of these models. As for the argument 
presented against the binary incursion model, namely that ``orbital periods set the 
time interval between arcs/rings, which is very regulated, conflicting with the 
fact that some arcs are intersecting, and time intervals are not exactly the same.",
it appears to me that a binary incursion might not
be as regular as the period of an undisturbed companion. Every time the companion
enters the AGB envelope in its elliptical orbit, 
it will lose some energy via friction and the orbit and period
will change as a result. Can this rescue the binary hypothesis with regards to ring production?

Venturing outside the volumes where the ring structures reside, large
faint ionized halos have been observed by Romano Corradi (see also Corradi et al. 2003),
by taking very deep images over 
large angular scales.
With few exceptions these are spherical structures, and like the rings, point to a 
spherical AGB mass-loss. For the few cases where multiple structures are encountered, 
the entire ejecta are approximately coeval, excluding that recurrent phenomena
might have taken place during the halo formation. Ionizing radiation, leaking from
mass-bounded, inner PNe gives a glimpse of the the mass-loss geometry preceding the sculpting
at the hand of the jet-like structures.

These observations, along with the
results presented in the ``outflows and jets, session II" by
Valent\'in Bujarrabal, where the momenta of several PNe and PPNe are shown to exceed the
radiation driving limit, appear to reinforce further the binary argument.
Personally, I have recently jumped on the band-wagon of the binary proponents. 
However there are some tough questions we are facing, some of which I outline below.

{\bf Where are the post-CE stars.} Having such detailed observations of
the close circumstellar quarters
of post-AGB stars, leaves me with a nagging question. 
If we are to admit to a sizable number of post-common envelope (CE) PNe to aid in the
explanation of PPNe morphologies, and if we
trust our (meager) understanding of CE ejections, we should be able to observe some peculiarities
in the circumstellar environments of those PPNe and PNe which descended from CE events.
CE events, are thought to be fast (time scales of one to a few 
decades) mass-losing events, leading to most or almost all of the AGB star envelope mass
(several tens of a solar mass)
and resulting in the winding in of the companion orbit (Sandquist et al. 1998; 
De Marco et al. 2003). The traumatized star should have a much reduced radius,
with a resulting increase in effective temperature and/or reduced luminosity,
but possibly be out of equilibrium. What are the expected excursions from the
AGB position on the HR diagram?

{\bf Too Many Binaries?}
If we decided that binary companions were necessary to eject PNe on more than 10\% of cases, 
the number of PNe in
the galaxy might be too large to be produced by the known (close) binary main sequence population.
Until recently we could count on about 10\% close binary central stars (e.g. Bond 2000),
leaving us in a position to wonder how single stars make PNe, but also safe
from the problem I am about to outline.
The central star radial velocity surveys of Don Pollacco and 
collaborators and Howard Bond and collaborators (see De Marco et al., these proceedings), 
show a high ($\sim$45\%) percentage of radial
velocity variables, possibly pointing to a similar fraction of post-CE
binaries.
According to Mathieu (1992) the main sequence binary fraction is of the order 60\%. However,
only $\sim$10\% have periods which are short enough to result in a CE event. If we assume that
of the 100 billions stars in the galaxy, only 10\% have lifetimes shorter than the age of the Universe,
and if we adopt a mean life for any such star of 10 billion years, and a mean life for a PN of
about 20\,000 years, then we expect to have only 2000 post-CE PN living in the Galaxy or $\sim$4000 PNe
in total. This is too few by a factor of five, since recent counts, which include visibility effects
point to $\sim$20\,000 PNe.
If the results of Pollacco and Bond don't go away, we might rejoice in having explained how PNe
form, but we will have a problem in explaining the number of binary central stars from a
population perspective.

{\bf In conclusion,} binaries might well be simpler to explain the large majority of PNe and,
according to recent surveys,
might well be present. However, we must reconcile their number with the main sequence binaries. 
Once it becomes clearer how many post CE PNe are out there, we should agree on a prototypical
CE model and its observational predictions. Noam Soker might say that no model can be accurate 
enough, but I think we must start from a square cow if we are to end with one with legs and
ears and a bell.

\end{document}